\documentclass[
 reprint,
amsmath,amssymb,
aps,
]{revtex4-2}
\usepackage{graphicx}
\usepackage{dcolumn}
\usepackage{bm}
\usepackage{lineno}
\usepackage{color}

\begin{document}

\preprint{APS/123-QED}

\title{Light-field dressing of  transient photo-excited states above $E_F$}

\author{Fei Wang${}^{1,\ast}$, Wanying Chen${}^{1,\ast}$, Changhua Bao${}^{1}$, Tianyun Lin${}^{1}$, Haoyuan Zhong${}^{1}$,\\ Hongyun Zhang${}^{1}$, Shuyun Zhou${}^{1,2,\dagger}$}

\affiliation{${}^{1}$State Key Laboratory of Low-Dimensional Quantum Physics and Department of Physics,Tsinghua University, Beijing 100084, People's Republic of China\\
${}^{2}$Frontier Science Center for Quantum Information, Beijing 100084, People's Republic of China}

\date{\today}
\begin{abstract}

Time-periodic light-field provides an emerging pathway for dynamically engineering quantum materials by forming hybrid states between photons and Bloch electrons. So far, experimental progress on  light-field dressed states has been mainly focused on the occupied states, however, it is unclear if the transient photo-excited states above the Fermi energy $E_F$ can also be dressed, leaving the dynamical interplay between photo-excitation and light-field dressing elusive. Here, we provide direct experimental evidence for light-field dressing of the transient photo-excited surface states above $E_F$, which exhibits distinct  dynamics with a delay response as compared to light-field dressed states below $E_F$. Our work reveals the dual roles of the pump pulse in both photo-excitation and light-field dressing, providing a more comprehensive picture with new insights on the light-induced manipulation of transient electronic states.

\end{abstract}

\maketitle

Three-dimensional topological insulators (TIs), which are characterized by insulating bulk states and topologically-protected surface state (SS) crossing the Fermi energy $E_F$, provide a fascinating platform for investigating light-matter interaction dynamics and ultrafast manipulation of quantum materials \cite{Sentef2021,ZhouNRP2021,Hsieh2017towards,marsi2018ultrafast}. The conical electronic structure of SS makes it easy to ``resonantly'' photo-excite electrons to  the unoccupied states above $E_F$ \cite{sobota2012ultrafast,wang2012measurement,sobota2014distinguishing,kuroda2016generation,Lanzara2016spin}, where the energy difference between the occupied and unoccupied states is provided by the pump photon energy. On the other hand, under strong light-matter interaction, the time-periodic light-field can also dress Bloch states inside the crystal, forming Floquet-Bloch states via virtual absorption or emission of photons \cite{oka2019floquet,rudner2020NRP}. 
Such electron-photon hybrid states can further lead to modification of the transient electronic structure \cite{Gedik2013,Gedik2016,zhou2023pseudospin,Huber2023build,Cavalleri2020light,ZhouBPPRL2023,beaulieu2021ultrafast}, optical properties \cite{Wang2014Stark,GedikOpticalStark2015,Hsieh2021nat,Anshul2024natmat,uchida2022diabatic,GhimireNP2023}, and tunneling current \cite{park2022steady} etc, thereby providing a dynamical control knob for engineering the material properties dubbed as Floquet engineering.

Time- and angle-resolved photoemission spectroscopy (TrARPES) is a powerful experimental technique for investigating Floquet engineering as well as photo-excited carrier dynamics \cite{sobota2021angle,ZhangNRMP22,boschini2023time}. Important progress such as light-induced engineering of the electronic structure have been reported in topological insulator \cite{Gedik2013,Gedik2016} and black phosphorus\cite{zhou2023pseudospin, ZhouBPPRL2023}. The formation dynamics of the Floquet states \cite{Huber2023build} have been reported in topological insulators. However, so far, most of the experimental progress has been focused on the occupied SS (occ-SS), while little is known about the light-field dressing of the unoccupied electronic states above $E_F$. For example, it is unclear if the states above $E_F$ can also be dressed by the light-field after being transiently populated by photo-excited carriers, and if yes, whether they exhibit different dynamics. Experimental investigation of the dynamical interplay between photo-excitation and light-field dressing can extend our knowledge on the fundamental physics of light-matter coupling under strong light-field.

Here, by performing TrARPES measurements on a  $p$-type (hole-doped) bismuth telluride (Bi$_2$Te$_3$) with mid-infrared (MIR) pumping at 220 meV photon energy [see schematic in Fig.~1(a)], we report light-induced sidebands of different electronic states, including the bulk valence band (BVB) and occ-SS, as well as the transiently photo-excited surface state (ex-SS) above $E_F$. More importantly, these sidebands exhibit distinct dynamics. In particular, light-field induced sidebands of BVB and occ-SS emerge simultaneously with the pump field and disappear before thermalization between the occ-SS and ex-SS is completed. In contrast, light-field induced sidebands of the transiently populated ex-SS show a delay response, reflecting the convoluted dynamics between photo-excitation and light-field dressing. Our work extends fundamental understanding to the physics of light-field dressed states of the ex-SS above $E_F$, and provides perspectives on engineering of such transiently-excited electronic states  above $E_F$ via two-color pumping TrARPES measurements.

The sample under investigation is a hole-doped Bi$_2$Te$_3$ topological insulator \cite{Huber2023build,kokh2014melt,chen2013tunable} with the Dirac point slightly below $E_F$ [Fig.~1(b)]. The doping is chosen such that the Fermi energy cuts through the SS, dividing it into the occ-SS below $E_F$, and unoccupied states above $E_F$ which become detectable in TrARPES measurements only after they are transiently populated by photo-excited carriers. The dynamical interplay between photo-excitation and light-field dressing of these surface states are the main focus of this work. 

\begin{figure*}[htbp]
	\includegraphics{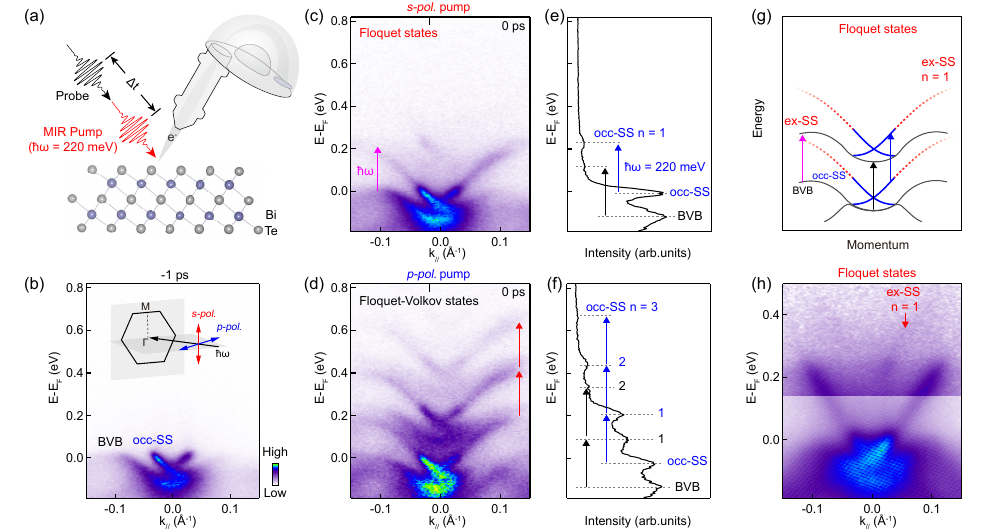}
	\caption{
	{(a)} Schematic for TrARPES with MIR pumping.
	{(b)} Dispersion image measured at $\Delta t = -1$ ps. The inset shows a sketch of the experimental geometry for the TrARPES setup.
	{(c)-(d)} TrARPES dispersion images measured at $\Delta t = 0$ ps for \textit{s-pol.} {(c)} and \textit{p-pol.} {(d)} ~pump polarizations at pump fluence of 1.0 mJ/cm$^2$.
	{(e)-(f)} EDCs at the dashed line marked in {(c)} and {(d)}. 
    {(g)} Schematic for  ex-SS and Floquet sidebands of occ-SS and BVB.
    {(h)} Similar data to {(c)} but at a higher pump fluence of 1.5 mJ/cm$^2$.
    }
    \label{Fig1}
\end{figure*}

Figures 1(c) and 1(d) show TrARPES snapshots measured at delay time $\Delta t = 0$ when the pump and probe pulses overlap, using two different pump polarizations. We note that light-fields with s-polarization (\textit{s-pol.}) and p-polarization (\textit{p-pol.}) can have different couplings to Bloch electrons inside the crystal and photo-emitted electrons in the vacuum, leading to pure Floquet-Bloch states and Floquet-Volkov states respectively \cite{Gedik2016}. For \textit{s-pol.} pump [Fig.~1(c)], a sideband displaced from the occ-SS by 220 meV - the pump photon energy, is clearly observed in the energy distribution curve (EDC) shown in Fig.~1(e), indicating the existence of Floquet-Bloch states [schematically illustrated in Fig.~1(g)]. In addition, photo-excited electrons are populated from the BVB into the SS above $E_F$ [pointed by pink arrow in Fig.~1(c)], allowing the detection of ex-SS  [red dashed curve in Fig.~1(g)] by TrARPES. Compared to \textit{s-pol.} pump, \textit{p-pol.} pump [Fig.~1(d)] shows a more complicated dispersion image with multiple sets of sidebands due to the interference between the Floquet and Volkov states. Here sidebands of occ-SS up to n = 3 order [marked by blue arrows in Fig.~1(f)], sidebands of BVB up to n = 2 order [indicated by black arrows in Fig.~1(f)] are clearly observed. Moreover, the ex-SS also shows clear sidebands at least up to n = 2 order [marked by red arrows in Fig.~1(d)]. More importantly, by using \textit{s-pol.} pump at a higher pump fluence, pure Floquet-Bloch states of the ex-SS are observed in Fig.~1(h). These data clearly show that not only the occ-SS and BVB, but also the transiently ex-SS is dressed by the light-field. Our results therefore provide direct experimental evidence that the pump pulse plays two roles, photo-exciting transient states and further dressing these states via the time-periodic light-field.

\begin{figure*}[htbp]
	\includegraphics{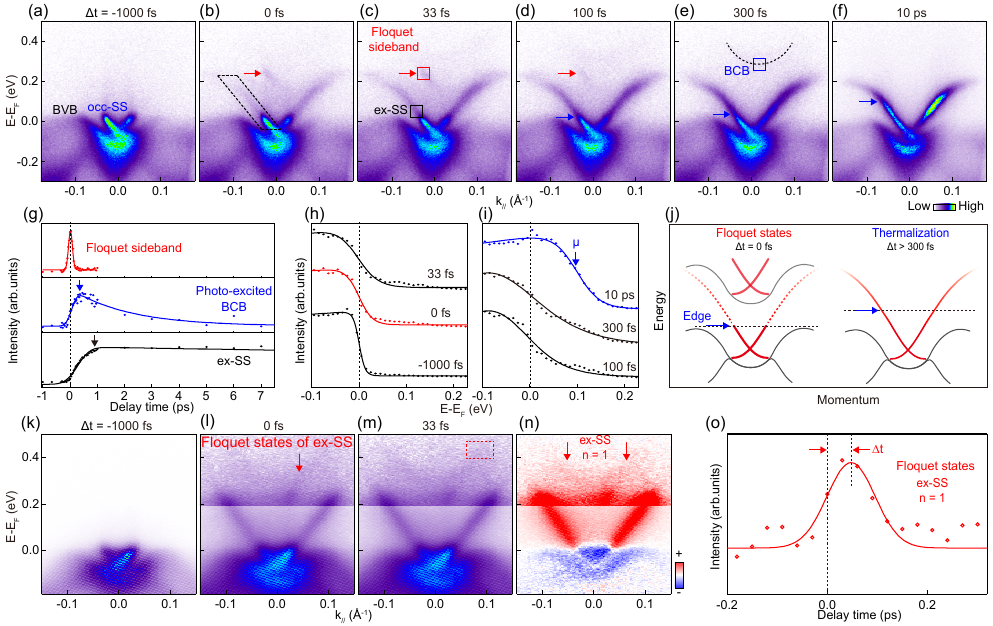}
	\caption{
     {(a)-(f)} TrARPES dispersion images at different delay times measured along the M-$\Gamma$-M  direction using \textit{s-pol.} pump. The pump photon energy is 220 meV and fluence is 1.0 mJ$\cdot$cm$^{-2}$. 
	 {(g)} Temporal evolution of the Floquet sideband, photo-excited BCB and ex-SS intensity.
     {(h)-(i)} EDCs at different delay times to show the thermalization between ex-SS and occ-SS.
     {(j)} Schematic for ex-SS and Floquet sidebands of BVB, occ-SS (left), and thermalization between ex-SS and occ-SS (right).
     {(k)-(m)} TrARPES dispersion images at different delay times measured along
     the K-$\Gamma$-K direction using \textit{s-pol.} pump. The pump photon energy is 220 meV and the fluence is 1.5 mJ$\cdot$cm$^{-2}$. 
     {(n)} Differential image obtained by subtracting data measured at $\Delta t$ = -1 ps from 0 ps.
     {(o)} Temporal evolution of the intensity for n = 1 Floquet sideband of ex-SS obtained by integrating the red dashed box in (m).
      }
	\label{Fig2}
\end{figure*}

To further reveal the dynamics of the Floquet states and photo-excited carriers, we show in Figs.~2(a)-2(f) snapshots of dispersion images measured along the M-$\Gamma$-M direction at different delay times using \textit{s-pol.} pump. The temporal evolution of the intensity for the Floquet sideband of the occ-SS, photo-excited BCB and ex-SS is shown in Fig.~2(g). The Floquet sideband of the occ-SS shows a sharp peak centered at $\Delta t = 0$, in agreement with what is expected for Floquet states. In contrast, the photo-excited BCB and ex-SS show a delay response with a much longer relaxation time. In particular, the BCB reaches maximum intensity at $\Delta t = 0.39$ ps [pointed by blue arrow in Fig.~2(g)], while the ex-SS reaches maximum intensity at a later time of $\Delta t = 0.95$ ps (pointed by black arrow), suggesting that there is relaxation of electrons from high-energy states to these states \cite{hajlaoui2012ultrafast,sanchez2017subpicosecond}, similar to the case of Bi$_2$Se$_3$ \cite{wang2012measurement,sobota2014distinguishing}. By fitting the relaxation dynamics with an exponential function, a lifetime of 2.1 ± 0.3 ps and 81 ± 12 ps is obtained for the BCB and ex-SS respectively. 
To further reveal the relaxation between the occ-SS and ex-SS, we show in Figs.~2(h) and 2(i) the integrated intensity of the SS [marked by dashed parallelogram in Fig.~2(b)] as a function of energy. The abrupt intensity jump, which corresponds to the Fermi edge at $\Delta t = -1000$ fs, becomes broader at $\Delta t = 0$ and later delay time, e.g. $\Delta t = 300$ fs, indicating a higher electronic temperature upon pumping. At even later delay time $\Delta t = 10$ ps, the width of the edge becomes narrower than that at $\Delta t = 300$ fs (Fig.~2(i) and Fig.~S1 in the Supplemental Material \cite{ref1}), suggesting a decrease of the electronic temperature. The above analysis shows that the Floquet states emerge near $\Delta t = 0$, and disappear before scattering or thermalization between the occ-SS and the ex-SS is completed, as schematically summarized in Fig.~2(j). Meanwhile, the chemical potential is clearly shifted by 0.1 eV with respect to $E_F$ due to surface photovoltaic effect \cite{hajlaoui2014tuning}. Here, the accumulation of transient electrons in the surface state above $E_F$ with a sufficiently long lifetime is useful for the formation of Floquet sidebands.

Figures 2(k)-2(m) show snapshots of dispersion images measured with high-purity \textit{s-pol.} pump (Figs.~S2-S3 in the Supplemental Material \cite{ref1} for more information about the geometry and the polarization) at a higher pump fluence. Figure 2(n) shows the differential dispersion image between $\Delta t$ = 0 ps and $\Delta t$ = -1 ps, where the Floquet sideband of ex-SS is clearly observed. Figure 2(o) shows the temporal evolution of the intensity for the first-order sideband of ex-SS, where the intensity maximum occurs at a later delay time of 46 ± 18 fs, suggesting that there is a dynamic interplay between photo-excitation and Floquet sideband formation.

\begin{figure*}[htbp]
	\includegraphics{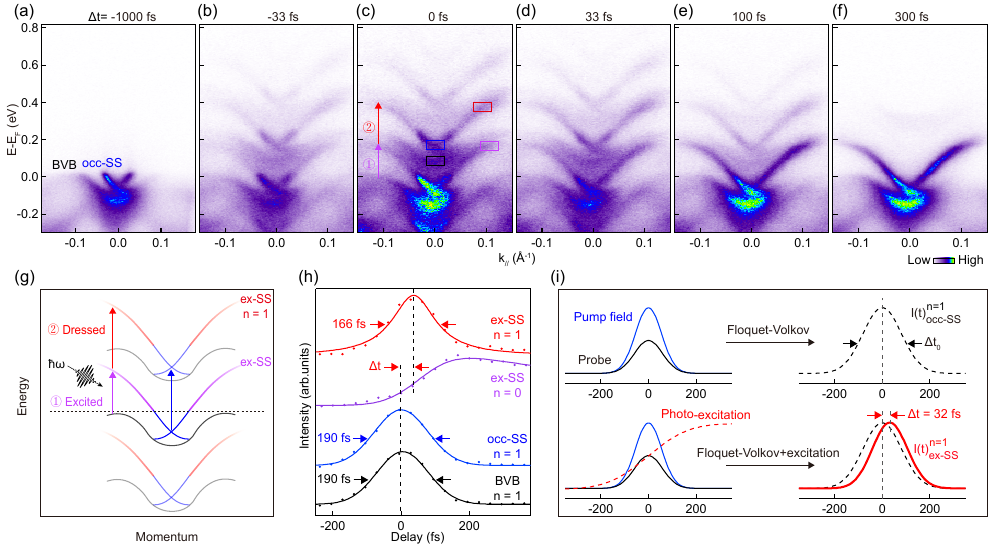}
  \caption{
    {(a)-(f)} TrARPES dispersion images measured at different delay times along M-$\Gamma$-M  direction with \textit{p-pol.} pump.
    {(g)} Schematic to illustrate the two roles of the pump pulse: photo-excitation and light-field dressing.
    {(h)} Temporal evolution of the intensity for n = 1 sidebands of BVB, occ-SS, ex-SS and original (n = 0) ex-SS (integrated over the black, blue, red and purple box in (c)).
    {(i)} The simulated temporal profile of light-field and n = 1 sidebands of occ-SS and ex-SS. 
    }
    \label{Fig3}
\end{figure*}

We further investigate the dynamical evolution of the light-field dressed ex-SS, and its comparison with light-field dressed occ-SS.  Here \textit{p-pol.} pump is used to enhance optical transitions into the ex-SS, which also results in a stronger intensity for light-field induced sidebands. Figures~3(a)-3(f) show snapshots of dispersion images measured at different delay times, where the pump pulse clearly excites electrons into the ex-SS, and the light-field further dresses the ex-SS [Fig.~3(g)]. To extract the duration of the Floquet-Volkov states, we show in Fig.~3(h) the temporal evolution of the intensity for the first-order sidebands for BVB, occ-SS and ex-SS. The comparison shows that the intensity maximum of the ex-SS sideband occurs at a later delay time by 35 $\pm$ 6 fs than the sidebands of BVB and occ-SS, spanning a shorter time window of 166 $\pm$ 6 fs compared to 190 $\pm$ 3 fs. We note that the extracted duration of the sidebands of the occupied states is determined by the TrARPES instrumental time resolution, namely, the convolution between the temporal profiles of the pump and probe pulses (blue and black curves in Fig.~3(i)), while the dynamics of ex-SS sideband is more complicated involving an additional photo-excitation process. The dynamical evolution of the ex-SS band [purple curve in Fig.~3(h)] shows that its rising edge is delayed from time zero by 43 $\pm$ 15 fs, which is similar to the n = 1 sideband whose intensity maximum occurs at a later delay time by 35 ± 6 fs. This suggests that the delay response of the n = 1 sideband is related to the occupation of electrons in the ex-SS above $E_F$. The latter involves scattering from high-energy electronic states into low-energy ex-SS after photo-excitation, which can be modeled by the convolution between a step function and an exponential function. After including these additional processes [red dashed curve in Fig.~3(i)] into the simulation, the simulated dynamics shows a delay response with 32 fs [red solid curve in Fig.~3(i)], which is in overall agreement with our experimental observation (Fig.~S4 in the Supplemental Material \cite{ref1}). We note that similar delay response of the photo-excited surface state is also observed in Bi$_{0.58}$Sb$_{1.42 }$Te$_3$and Bi$_2$Se$_3$ with different doping (Fig.~S5 in the Supplemental Material \cite{ref1}), where the delay time slightly varies due to the different dynamics of the transient photo-excited states. More importantly, a comparison of experimental results on these different samples (Fig.~S6 in the Supplemental Material \cite{ref1}) shows that the sidebands and the crossing points are much better resolved in Bi$_{0.58}$Sb$_{1.42}$Te$_3$ compared to Bi$_2$Se$_3$. This suggests that photo-excitation by the pump can play the role of electron doping, while having the advantage of reducing scattering due to the little occupation of BCB and facilitating the detection of light-field dressed states.

\begin{figure*}[htbp]
	\includegraphics{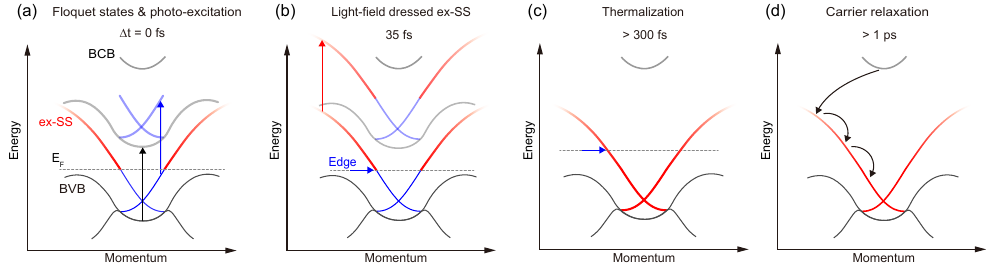}
	\caption{
	{(a)-(d)} Schematic to illustrate the dynamics at different delay times, which involve light-field dressed states and photo-excitation (a), light-field dressed ex-SS (b), thermalization between ex-SS and occ-SS (c), and carrier relaxation from BCB to ex-SS (d).
 	}\label{Fig4}
\end{figure*}

Combining the above experimental results, the dynamics at different delay times can be summarized by schematic illustrations in Figs.~4(a)-4(c). At time zero, the light-field of the pump pulse dresses the occupied electronic states, meanwhile the pump pulse also excites electrons into the BCB and ex-SS [Fig.~4(a)]. These transiently excited states are also dressed by the light-field, and their sidebands emerge at a slightly later time [Fig.~4(b)]. The sidebands disappear when the pump and probe pulses do not overlap, meanwhile electrons in the ex-SS and occ-SS continue to thermalize, leading to a shift of the edge above $E_F$ [Fig.~4(c)]. 
Moreover, by analyzing the dynamics over a larger temporal window (Fig.~S7 in the Supplemental Material \cite{ref1}), we reveal that photo-excited carriers in the BCB relax to ex-SS, and eventually relax to the equilibrium state via intra-band scattering [Fig.~4(d)]. These results together provide a more complete picture on the dynamics including the light-field dressing and relaxation of photo-excited carriers. These results provide new insights into the physics of light-field dressing of photo-excited states above $E_F$, and suggest potential applications of two-color pumping TrARPES in the Floquet engineering of graphene, transition metal dichalcogenides (TMDCs) etc. For example, for monolayer transition metal dichalcogenide, a circular polarized pump with photon energy above the band gap can be first used to selectively excite electrons to the conduction band at specific momentum valley (K or K'), while a second pump with a lower pump photon energy and a stronger light-field can further dress the transiently excited valley-polarized conduction band electrons. Therefore, two-color pumping TrARPES measurements can combine the advantages of dynamical photo-doping and light-field dressing, providing richer tunabilities.

In summary, TrARPES measurements on $p$-type Bi$_2$Te$_3$ with MIR pumping allow to provide direct experimental evidence for the light-field dressing of the transiently excited states above $E_F$, and elucidate the rich interplay between photo-excitation and light-field dressing. These experimental results provide a more complete picture on the physics of light-induced manipulation of quantum materials with new insights on the physics of light-field dressed transiently excited states. For example, by two-color pumping TrARPES, the transiently-occupied states can be manipulated with even richer tunabilities by combining the advantages of dynamical photo-doping and light-field dressing.

This work is supported by the National Natural Science Foundation of China (Grant No.~12234011, 52388201, 92250305, 12327805, 11725418, 11427903), National Key R \& D Program of China (Grant No.~2021YFA1400100), and New Cornerstone Science Foundation through the XPLORER PRIZE.

$^\ast$These authors contributed equally to this work

$^\dagger$Corresponding author: syzhou@mail.tsinghua.edu.cn


\begin{thebibliography}{34}%
  \makeatletter
  \providecommand \@ifxundefined [1]{%
   \@ifx{#1\undefined}
  }%
  \providecommand \@ifnum [1]{%
   \ifnum #1\expandafter \@firstoftwo
   \else \expandafter \@secondoftwo
   \fi
  }%
  \providecommand \@ifx [1]{%
   \ifx #1\expandafter \@firstoftwo
   \else \expandafter \@secondoftwo
   \fi
  }%
  \providecommand \natexlab [1]{#1}%
  \providecommand \enquote  [1]{``#1''}%
  \providecommand \bibnamefont  [1]{#1}%
  \providecommand \bibfnamefont [1]{#1}%
  \providecommand \citenamefont [1]{#1}%
  \providecommand \href@noop [0]{\@secondoftwo}%
  \providecommand \href [0]{\begingroup \@sanitize@url \@href}%
  \providecommand \@href[1]{\@@startlink{#1}\@@href}%
  \providecommand \@@href[1]{\endgroup#1\@@endlink}%
  \providecommand \@sanitize@url [0]{\catcode `\\12\catcode `\$12\catcode
    `\&12\catcode `\#12\catcode `\^12\catcode `\_12\catcode `\%12\relax}%
  \providecommand \@@startlink[1]{}%
  \providecommand \@@endlink[0]{}%
  \providecommand \url  [0]{\begingroup\@sanitize@url \@url }%
  \providecommand \@url [1]{\endgroup\@href {#1}{\urlprefix }}%
  \providecommand \urlprefix  [0]{URL }%
  \providecommand \Eprint [0]{\href }%
  \providecommand \doibase [0]{https://doi.org/}%
  \providecommand \selectlanguage [0]{\@gobble}%
  \providecommand \bibinfo  [0]{\@secondoftwo}%
  \providecommand \bibfield  [0]{\@secondoftwo}%
  \providecommand \translation [1]{[#1]}%
  \providecommand \BibitemOpen [0]{}%
  \providecommand \bibitemStop [0]{}%
  \providecommand \bibitemNoStop [0]{.\EOS\space}%
  \providecommand \EOS [0]{\spacefactor3000\relax}%
  \providecommand \BibitemShut  [1]{\csname bibitem#1\endcsname}%
  \let\auto@bib@innerbib\@empty
  \bibitem [{\citenamefont {de~la Torre}\ \emph {et~al.}(2021)\citenamefont
    {de~la Torre}, \citenamefont {Kennes}, \citenamefont {Claassen},
    \citenamefont {Gerber}, \citenamefont {McIver},\ and\ \citenamefont
    {Sentef}}]{Sentef2021}%
    \BibitemOpen
    \bibfield  {author} {\bibinfo {author} {\bibfnamefont {A.}~\bibnamefont
    {de~la Torre}}, \bibinfo {author} {\bibfnamefont {D.~M.}\ \bibnamefont
    {Kennes}}, \bibinfo {author} {\bibfnamefont {M.}~\bibnamefont {Claassen}},
    \bibinfo {author} {\bibfnamefont {S.}~\bibnamefont {Gerber}}, \bibinfo
    {author} {\bibfnamefont {J.~W.}\ \bibnamefont {McIver}},\ and\ \bibinfo
    {author} {\bibfnamefont {M.~A.}\ \bibnamefont {Sentef}},\ }\bibfield  {title}
    {\bibinfo {title} {{Colloquium: Nonthermal pathways to ultrafast control in
    quantum materials}},\ }\href {https://doi.org/10.1103/RevModPhys.93.041002}
    {\bibfield  {journal} {\bibinfo  {journal} {Rev. Mod. Phys.}\ }\textbf
    {\bibinfo {volume} {93}},\ \bibinfo {pages} {041002} (\bibinfo {year}
    {2021})}\BibitemShut {NoStop}%
  \bibitem [{\citenamefont {Bao}\ \emph {et~al.}(2021)\citenamefont {Bao},
    \citenamefont {Tang}, \citenamefont {Sun},\ and\ \citenamefont
    {Zhou}}]{ZhouNRP2021}%
    \BibitemOpen
    \bibfield  {author} {\bibinfo {author} {\bibfnamefont {C.}~\bibnamefont
    {Bao}}, \bibinfo {author} {\bibfnamefont {P.}~\bibnamefont {Tang}}, \bibinfo
    {author} {\bibfnamefont {D.}~\bibnamefont {Sun}},\ and\ \bibinfo {author}
    {\bibfnamefont {S.}~\bibnamefont {Zhou}},\ }\bibfield  {title} {\bibinfo
    {title} {{Light-induced emergent phenomena in 2D materials and topological
    materials}},\ }\href@noop {} {\bibfield  {journal} {\bibinfo  {journal} {Nat.
    Rev. Phys.}\ }\textbf {\bibinfo {volume} {4}},\ \bibinfo {pages} {33}
    (\bibinfo {year} {2021})}\BibitemShut {NoStop}%
  \bibitem [{\citenamefont {Basov}\ \emph {et~al.}(2017)\citenamefont {Basov},
    \citenamefont {Averitt},\ and\ \citenamefont {Hsieh}}]{Hsieh2017towards}%
    \BibitemOpen
    \bibfield  {author} {\bibinfo {author} {\bibfnamefont {D.}~\bibnamefont
    {Basov}}, \bibinfo {author} {\bibfnamefont {R.}~\bibnamefont {Averitt}},\
    and\ \bibinfo {author} {\bibfnamefont {D.}~\bibnamefont {Hsieh}},\ }\bibfield
     {title} {\bibinfo {title} {Towards properties on demand in quantum
    materials},\ }\href@noop {} {\bibfield  {journal} {\bibinfo  {journal} {Nat.
    Mater.}\ }\textbf {\bibinfo {volume} {16}},\ \bibinfo {pages} {1077}
    (\bibinfo {year} {2017})}\BibitemShut {NoStop}%
  \bibitem [{\citenamefont {Marsi}(2018)}]{marsi2018ultrafast}%
    \BibitemOpen
    \bibfield  {author} {\bibinfo {author} {\bibfnamefont {M.}~\bibnamefont
    {Marsi}},\ }\bibfield  {title} {\bibinfo {title} {{Ultrafast electron
    dynamics in topological materials}},\ }\href@noop {} {\bibfield  {journal}
    {\bibinfo  {journal} {Phys. Status Solidi (RRL)}\ }\textbf {\bibinfo {volume}
    {12}},\ \bibinfo {pages} {1800228} (\bibinfo {year} {2018})}\BibitemShut
    {NoStop}%
  \bibitem [{\citenamefont {Sobota}\ \emph {et~al.}(2012)\citenamefont {Sobota},
    \citenamefont {Yang}, \citenamefont {Analytis}, \citenamefont {Chen},
    \citenamefont {Fisher}, \citenamefont {Kirchmann},\ and\ \citenamefont
    {Shen}}]{sobota2012ultrafast}%
    \BibitemOpen
    \bibfield  {author} {\bibinfo {author} {\bibfnamefont {J.~A.}\ \bibnamefont
    {Sobota}}, \bibinfo {author} {\bibfnamefont {S.}~\bibnamefont {Yang}},
    \bibinfo {author} {\bibfnamefont {J.~G.}\ \bibnamefont {Analytis}}, \bibinfo
    {author} {\bibfnamefont {Y.}~\bibnamefont {Chen}}, \bibinfo {author}
    {\bibfnamefont {I.~R.}\ \bibnamefont {Fisher}}, \bibinfo {author}
    {\bibfnamefont {P.~S.}\ \bibnamefont {Kirchmann}},\ and\ \bibinfo {author}
    {\bibfnamefont {Z.-X.}\ \bibnamefont {Shen}},\ }\bibfield  {title} {\bibinfo
    {title} {{Ultrafast optical excitation of a persistent surface-state
    population in the topological insulator Bi$_2$Se$_3$}},\ }\href@noop {}
    {\bibfield  {journal} {\bibinfo  {journal} {Phys. Rev. Lett.}\ }\textbf
    {\bibinfo {volume} {108}},\ \bibinfo {pages} {117403} (\bibinfo {year}
    {2012})}\BibitemShut {NoStop}%
  \bibitem [{\citenamefont {Wang}\ \emph {et~al.}(2012)\citenamefont {Wang},
    \citenamefont {Hsieh}, \citenamefont {Sie}, \citenamefont {Steinberg},
    \citenamefont {Gardner}, \citenamefont {Lee}, \citenamefont
    {Jarillo-Herrero},\ and\ \citenamefont {Gedik}}]{wang2012measurement}%
    \BibitemOpen
    \bibfield  {author} {\bibinfo {author} {\bibfnamefont {Y.}~\bibnamefont
    {Wang}}, \bibinfo {author} {\bibfnamefont {D.}~\bibnamefont {Hsieh}},
    \bibinfo {author} {\bibfnamefont {E.}~\bibnamefont {Sie}}, \bibinfo {author}
    {\bibfnamefont {H.}~\bibnamefont {Steinberg}}, \bibinfo {author}
    {\bibfnamefont {D.}~\bibnamefont {Gardner}}, \bibinfo {author} {\bibfnamefont
    {Y.}~\bibnamefont {Lee}}, \bibinfo {author} {\bibfnamefont {P.}~\bibnamefont
    {Jarillo-Herrero}},\ and\ \bibinfo {author} {\bibfnamefont {N.}~\bibnamefont
    {Gedik}},\ }\bibfield  {title} {\bibinfo {title} {{Measurement of intrinsic
    Dirac fermion cooling on the surface of the topological insulator
    Bi$_2$Se$_3$ using time-resolved and angle-resolved photoemission
    spectroscopy}},\ }\href@noop {} {\bibfield  {journal} {\bibinfo  {journal}
    {Phys. Rev. Lett.}\ }\textbf {\bibinfo {volume} {109}},\ \bibinfo {pages}
    {127401} (\bibinfo {year} {2012})}\BibitemShut {NoStop}%
  \bibitem [{\citenamefont {Sobota}\ \emph {et~al.}(2014)\citenamefont {Sobota},
    \citenamefont {Yang}, \citenamefont {Leuenberger}, \citenamefont {Kemper},
    \citenamefont {Analytis}, \citenamefont {Fisher}, \citenamefont {Kirchmann},
    \citenamefont {Devereaux},\ and\ \citenamefont
    {Shen}}]{sobota2014distinguishing}%
    \BibitemOpen
    \bibfield  {author} {\bibinfo {author} {\bibfnamefont {J.~A.}\ \bibnamefont
    {Sobota}}, \bibinfo {author} {\bibfnamefont {S.-L.}\ \bibnamefont {Yang}},
    \bibinfo {author} {\bibfnamefont {D.}~\bibnamefont {Leuenberger}}, \bibinfo
    {author} {\bibfnamefont {A.~F.}\ \bibnamefont {Kemper}}, \bibinfo {author}
    {\bibfnamefont {J.~G.}\ \bibnamefont {Analytis}}, \bibinfo {author}
    {\bibfnamefont {I.~R.}\ \bibnamefont {Fisher}}, \bibinfo {author}
    {\bibfnamefont {P.~S.}\ \bibnamefont {Kirchmann}}, \bibinfo {author}
    {\bibfnamefont {T.~P.}\ \bibnamefont {Devereaux}},\ and\ \bibinfo {author}
    {\bibfnamefont {Z.-X.}\ \bibnamefont {Shen}},\ }\bibfield  {title} {\bibinfo
    {title} {{Distinguishing bulk and surface electron-phonon coupling in the
    topological insulator Bi$_2$Se$_3$ using time-resolved photoemission
    spectroscopy}},\ }\href@noop {} {\bibfield  {journal} {\bibinfo  {journal}
    {Phys. Rev. Lett.}\ }\textbf {\bibinfo {volume} {113}},\ \bibinfo {pages}
    {157401} (\bibinfo {year} {2014})}\BibitemShut {NoStop}%
  \bibitem [{\citenamefont {Kuroda}\ \emph {et~al.}(2016)\citenamefont {Kuroda},
    \citenamefont {Reimann}, \citenamefont {G{\"u}dde},\ and\ \citenamefont
    {H{\"o}fer}}]{kuroda2016generation}%
    \BibitemOpen
    \bibfield  {author} {\bibinfo {author} {\bibfnamefont {K.}~\bibnamefont
    {Kuroda}}, \bibinfo {author} {\bibfnamefont {J.}~\bibnamefont {Reimann}},
    \bibinfo {author} {\bibfnamefont {J.}~\bibnamefont {G{\"u}dde}},\ and\
    \bibinfo {author} {\bibfnamefont {U.}~\bibnamefont {H{\"o}fer}},\ }\bibfield
    {title} {\bibinfo {title} {{Generation of transient photocurrents in the
    topological surface state of Sb$_2$Te$_3$ by direct optical excitation with
    midinfrared pulses}},\ }\href@noop {} {\bibfield  {journal} {\bibinfo
    {journal} {Phys. Rev. Lett.}\ }\textbf {\bibinfo {volume} {116}},\ \bibinfo
    {pages} {076801} (\bibinfo {year} {2016})}\BibitemShut {NoStop}%
  \bibitem [{\citenamefont {Jozwiak}\ \emph {et~al.}(2016)\citenamefont
    {Jozwiak}, \citenamefont {Sobota}, \citenamefont {Gotlieb}, \citenamefont
    {Kemper}, \citenamefont {Rotundu}, \citenamefont {Birgeneau}, \citenamefont
    {Hussain}, \citenamefont {Lee}, \citenamefont {Shen},\ and\ \citenamefont
    {Lanzara}}]{Lanzara2016spin}%
    \BibitemOpen
    \bibfield  {author} {\bibinfo {author} {\bibfnamefont {C.}~\bibnamefont
    {Jozwiak}}, \bibinfo {author} {\bibfnamefont {J.~A.}\ \bibnamefont {Sobota}},
    \bibinfo {author} {\bibfnamefont {K.}~\bibnamefont {Gotlieb}}, \bibinfo
    {author} {\bibfnamefont {A.~F.}\ \bibnamefont {Kemper}}, \bibinfo {author}
    {\bibfnamefont {C.~R.}\ \bibnamefont {Rotundu}}, \bibinfo {author}
    {\bibfnamefont {R.~J.}\ \bibnamefont {Birgeneau}}, \bibinfo {author}
    {\bibfnamefont {Z.}~\bibnamefont {Hussain}}, \bibinfo {author} {\bibfnamefont
    {D.-H.}\ \bibnamefont {Lee}}, \bibinfo {author} {\bibfnamefont {Z.-X.}\
    \bibnamefont {Shen}},\ and\ \bibinfo {author} {\bibfnamefont
    {A.}~\bibnamefont {Lanzara}},\ }\bibfield  {title} {\bibinfo {title}
    {Spin-polarized surface resonances accompanying topological surface state
    formation},\ }\href@noop {} {\bibfield  {journal} {\bibinfo  {journal} {Nat.
    Commun.}\ }\textbf {\bibinfo {volume} {7}},\ \bibinfo {pages} {13143}
    (\bibinfo {year} {2016})}\BibitemShut {NoStop}%
  \bibitem [{\citenamefont {Oka}\ and\ \citenamefont
    {Kitamura}(2019)}]{oka2019floquet}%
    \BibitemOpen
    \bibfield  {author} {\bibinfo {author} {\bibfnamefont {T.}~\bibnamefont
    {Oka}}\ and\ \bibinfo {author} {\bibfnamefont {S.}~\bibnamefont {Kitamura}},\
    }\bibfield  {title} {\bibinfo {title} {{Floquet engineering of quantum
    materials}},\ }\href@noop {} {\bibfield  {journal} {\bibinfo  {journal}
    {Annu. Rev. Condens. Matter Phys.}\ }\textbf {\bibinfo {volume} {10}},\
    \bibinfo {pages} {387} (\bibinfo {year} {2019})}\BibitemShut {NoStop}%
  \bibitem [{\citenamefont {Rudner}\ and\ \citenamefont
    {Lindner}(2020)}]{rudner2020NRP}%
    \BibitemOpen
    \bibfield  {author} {\bibinfo {author} {\bibfnamefont {M.~S.}\ \bibnamefont
    {Rudner}}\ and\ \bibinfo {author} {\bibfnamefont {N.~H.}\ \bibnamefont
    {Lindner}},\ }\bibfield  {title} {\bibinfo {title} {{Band structure
    engineering and non-equilibrium dynamics in Floquet topological
    insulators}},\ }\href@noop {} {\bibfield  {journal} {\bibinfo  {journal}
    {Nat. Rev. Phys.}\ }\textbf {\bibinfo {volume} {2}},\ \bibinfo {pages} {229}
    (\bibinfo {year} {2020})}\BibitemShut {NoStop}%
  \bibitem [{\citenamefont {Wang}\ \emph {et~al.}(2013)\citenamefont {Wang},
    \citenamefont {Steinberg}, \citenamefont {Jarillo-Herrero},\ and\
    \citenamefont {Gedik}}]{Gedik2013}%
    \BibitemOpen
    \bibfield  {author} {\bibinfo {author} {\bibfnamefont {Y.}~\bibnamefont
    {Wang}}, \bibinfo {author} {\bibfnamefont {H.}~\bibnamefont {Steinberg}},
    \bibinfo {author} {\bibfnamefont {P.}~\bibnamefont {Jarillo-Herrero}},\ and\
    \bibinfo {author} {\bibfnamefont {N.}~\bibnamefont {Gedik}},\ }\bibfield
    {title} {\bibinfo {title} {{Observation of Floquet-Bloch states on the
    surface of a topological insulator}},\ }\href@noop {} {\bibfield  {journal}
    {\bibinfo  {journal} {Science}\ }\textbf {\bibinfo {volume} {342}},\ \bibinfo
    {pages} {453} (\bibinfo {year} {2013})}\BibitemShut {NoStop}%
  \bibitem [{\citenamefont {Mahmood}\ \emph {et~al.}(2016)\citenamefont
    {Mahmood}, \citenamefont {Chan}, \citenamefont {Alpichshev}, \citenamefont
    {Gardner}, \citenamefont {Lee}, \citenamefont {Lee},\ and\ \citenamefont
    {Gedik}}]{Gedik2016}%
    \BibitemOpen
    \bibfield  {author} {\bibinfo {author} {\bibfnamefont {F.}~\bibnamefont
    {Mahmood}}, \bibinfo {author} {\bibfnamefont {C.-K.}\ \bibnamefont {Chan}},
    \bibinfo {author} {\bibfnamefont {Z.}~\bibnamefont {Alpichshev}}, \bibinfo
    {author} {\bibfnamefont {D.}~\bibnamefont {Gardner}}, \bibinfo {author}
    {\bibfnamefont {Y.}~\bibnamefont {Lee}}, \bibinfo {author} {\bibfnamefont
    {P.~A.}\ \bibnamefont {Lee}},\ and\ \bibinfo {author} {\bibfnamefont
    {N.}~\bibnamefont {Gedik}},\ }\bibfield  {title} {\bibinfo {title}
    {{Selective scattering between Floquet-Bloch and Volkov states in a
    topological insulator}},\ }\href@noop {} {\bibfield  {journal} {\bibinfo
    {journal} {Nat. Phys.}\ }\textbf {\bibinfo {volume} {12}},\ \bibinfo {pages}
    {306} (\bibinfo {year} {2016})}\BibitemShut {NoStop}%
  \bibitem [{\citenamefont {Zhou}\ \emph
    {et~al.}(2023{\natexlab{a}})\citenamefont {Zhou}, \citenamefont {Bao},
    \citenamefont {Fan}, \citenamefont {Zhou}, \citenamefont {Gao}, \citenamefont
    {Zhong}, \citenamefont {Lin}, \citenamefont {Liu}, \citenamefont {Yu},
    \citenamefont {Tang} \emph {et~al.}}]{zhou2023pseudospin}%
    \BibitemOpen
    \bibfield  {author} {\bibinfo {author} {\bibfnamefont {S.}~\bibnamefont
    {Zhou}}, \bibinfo {author} {\bibfnamefont {C.}~\bibnamefont {Bao}}, \bibinfo
    {author} {\bibfnamefont {B.}~\bibnamefont {Fan}}, \bibinfo {author}
    {\bibfnamefont {H.}~\bibnamefont {Zhou}}, \bibinfo {author} {\bibfnamefont
    {Q.}~\bibnamefont {Gao}}, \bibinfo {author} {\bibfnamefont {H.}~\bibnamefont
    {Zhong}}, \bibinfo {author} {\bibfnamefont {T.}~\bibnamefont {Lin}}, \bibinfo
    {author} {\bibfnamefont {H.}~\bibnamefont {Liu}}, \bibinfo {author}
    {\bibfnamefont {P.}~\bibnamefont {Yu}}, \bibinfo {author} {\bibfnamefont
    {P.}~\bibnamefont {Tang}}, \emph {et~al.},\ }\bibfield  {title} {\bibinfo
    {title} {{Pseudospin-selective Floquet band engineering in black
    phosphorus}},\ }\href@noop {} {\bibfield  {journal} {\bibinfo  {journal}
    {Nature}\ }\textbf {\bibinfo {volume} {614}},\ \bibinfo {pages} {75}
    (\bibinfo {year} {2023}{\natexlab{a}})}\BibitemShut {NoStop}%
  \bibitem [{\citenamefont {Ito}\ \emph {et~al.}(2023)\citenamefont {Ito},
    \citenamefont {Sch{\"u}ler}, \citenamefont {Meierhofer}, \citenamefont
    {Schlauderer}, \citenamefont {Freudenstein}, \citenamefont {Reimann},
    \citenamefont {Afanasiev}, \citenamefont {Kokh}, \citenamefont
    {Tereshchenko}, \citenamefont {G{\"u}dde} \emph {et~al.}}]{Huber2023build}%
    \BibitemOpen
    \bibfield  {author} {\bibinfo {author} {\bibfnamefont {S.}~\bibnamefont
    {Ito}}, \bibinfo {author} {\bibfnamefont {M.}~\bibnamefont {Sch{\"u}ler}},
    \bibinfo {author} {\bibfnamefont {M.}~\bibnamefont {Meierhofer}}, \bibinfo
    {author} {\bibfnamefont {S.}~\bibnamefont {Schlauderer}}, \bibinfo {author}
    {\bibfnamefont {J.}~\bibnamefont {Freudenstein}}, \bibinfo {author}
    {\bibfnamefont {J.}~\bibnamefont {Reimann}}, \bibinfo {author} {\bibfnamefont
    {D.}~\bibnamefont {Afanasiev}}, \bibinfo {author} {\bibfnamefont
    {K.}~\bibnamefont {Kokh}}, \bibinfo {author} {\bibfnamefont {O.}~\bibnamefont
    {Tereshchenko}}, \bibinfo {author} {\bibfnamefont {J.}~\bibnamefont
    {G{\"u}dde}}, \emph {et~al.},\ }\bibfield  {title} {\bibinfo {title}
    {{Build-up and dephasing of Floquet--Bloch bands on subcycle timescales}},\
    }\href@noop {} {\bibfield  {journal} {\bibinfo  {journal} {Nature}\ }\textbf
    {\bibinfo {volume} {616}},\ \bibinfo {pages} {696} (\bibinfo {year}
    {2023})}\BibitemShut {NoStop}%
  \bibitem [{\citenamefont {McIver}\ \emph {et~al.}(2020)\citenamefont {McIver},
    \citenamefont {Schulte}, \citenamefont {Stein}, \citenamefont {Matsuyama},
    \citenamefont {Jotzu}, \citenamefont {Meier},\ and\ \citenamefont
    {Cavalleri}}]{Cavalleri2020light}%
    \BibitemOpen
    \bibfield  {author} {\bibinfo {author} {\bibfnamefont {J.~W.}\ \bibnamefont
    {McIver}}, \bibinfo {author} {\bibfnamefont {B.}~\bibnamefont {Schulte}},
    \bibinfo {author} {\bibfnamefont {F.-U.}\ \bibnamefont {Stein}}, \bibinfo
    {author} {\bibfnamefont {T.}~\bibnamefont {Matsuyama}}, \bibinfo {author}
    {\bibfnamefont {G.}~\bibnamefont {Jotzu}}, \bibinfo {author} {\bibfnamefont
    {G.}~\bibnamefont {Meier}},\ and\ \bibinfo {author} {\bibfnamefont
    {A.}~\bibnamefont {Cavalleri}},\ }\bibfield  {title} {\bibinfo {title}
    {{Light-induced anomalous Hall effect in graphene}},\ }\href@noop {}
    {\bibfield  {journal} {\bibinfo  {journal} {Nat. Phys.}\ }\textbf {\bibinfo
    {volume} {16}},\ \bibinfo {pages} {38} (\bibinfo {year} {2020})}\BibitemShut
    {NoStop}%
  \bibitem [{\citenamefont {Zhou}\ \emph
    {et~al.}(2023{\natexlab{b}})\citenamefont {Zhou}, \citenamefont {Bao},
    \citenamefont {Fan}, \citenamefont {Wang}, \citenamefont {Zhong},
    \citenamefont {Zhang}, \citenamefont {Tang}, \citenamefont {Duan},\ and\
    \citenamefont {Zhou}}]{ZhouBPPRL2023}%
    \BibitemOpen
    \bibfield  {author} {\bibinfo {author} {\bibfnamefont {S.}~\bibnamefont
    {Zhou}}, \bibinfo {author} {\bibfnamefont {C.}~\bibnamefont {Bao}}, \bibinfo
    {author} {\bibfnamefont {B.}~\bibnamefont {Fan}}, \bibinfo {author}
    {\bibfnamefont {F.}~\bibnamefont {Wang}}, \bibinfo {author} {\bibfnamefont
    {H.}~\bibnamefont {Zhong}}, \bibinfo {author} {\bibfnamefont
    {H.}~\bibnamefont {Zhang}}, \bibinfo {author} {\bibfnamefont
    {P.}~\bibnamefont {Tang}}, \bibinfo {author} {\bibfnamefont {W.}~\bibnamefont
    {Duan}},\ and\ \bibinfo {author} {\bibfnamefont {S.}~\bibnamefont {Zhou}},\
    }\bibfield  {title} {\bibinfo {title} {{Floquet Engineering of Black
    Phosphorus upon Below-Gap Pumping}},\ }\href@noop {} {\bibfield  {journal}
    {\bibinfo  {journal} {Phys. Rev. Lett.}\ }\textbf {\bibinfo {volume} {131}},\
    \bibinfo {pages} {116401} (\bibinfo {year} {2023}{\natexlab{b}})}\BibitemShut
    {NoStop}%
  \bibitem [{\citenamefont {Beaulieu}\ \emph {et~al.}(2021)\citenamefont
    {Beaulieu}, \citenamefont {Dong}, \citenamefont {Tancogne-Dejean},
    \citenamefont {Dendzik}, \citenamefont {Pincelli}, \citenamefont {Maklar},
    \citenamefont {Xian}, \citenamefont {Sentef}, \citenamefont {Wolf},
    \citenamefont {Rubio} \emph {et~al.}}]{beaulieu2021ultrafast}%
    \BibitemOpen
    \bibfield  {author} {\bibinfo {author} {\bibfnamefont {S.}~\bibnamefont
    {Beaulieu}}, \bibinfo {author} {\bibfnamefont {S.}~\bibnamefont {Dong}},
    \bibinfo {author} {\bibfnamefont {N.}~\bibnamefont {Tancogne-Dejean}},
    \bibinfo {author} {\bibfnamefont {M.}~\bibnamefont {Dendzik}}, \bibinfo
    {author} {\bibfnamefont {T.}~\bibnamefont {Pincelli}}, \bibinfo {author}
    {\bibfnamefont {J.}~\bibnamefont {Maklar}}, \bibinfo {author} {\bibfnamefont
    {R.~P.}\ \bibnamefont {Xian}}, \bibinfo {author} {\bibfnamefont {M.~A.}\
    \bibnamefont {Sentef}}, \bibinfo {author} {\bibfnamefont {M.}~\bibnamefont
    {Wolf}}, \bibinfo {author} {\bibfnamefont {A.}~\bibnamefont {Rubio}}, \emph
    {et~al.},\ }\bibfield  {title} {\bibinfo {title} {{Ultrafast dynamical
    Lifshitz transition}},\ }\href@noop {} {\bibfield  {journal} {\bibinfo
    {journal} {Sci. Adv.}\ }\textbf {\bibinfo {volume} {7}},\ \bibinfo {pages}
    {eabd9275} (\bibinfo {year} {2021})}\BibitemShut {NoStop}%
  \bibitem [{\citenamefont {Kim}\ \emph {et~al.}(2014)\citenamefont {Kim},
    \citenamefont {Hong}, \citenamefont {Jin}, \citenamefont {Shi}, \citenamefont
    {Chang}, \citenamefont {Chiu}, \citenamefont {Li},\ and\ \citenamefont
    {Wang}}]{Wang2014Stark}%
    \BibitemOpen
    \bibfield  {author} {\bibinfo {author} {\bibfnamefont {J.}~\bibnamefont
    {Kim}}, \bibinfo {author} {\bibfnamefont {X.}~\bibnamefont {Hong}}, \bibinfo
    {author} {\bibfnamefont {C.}~\bibnamefont {Jin}}, \bibinfo {author}
    {\bibfnamefont {S.-F.}\ \bibnamefont {Shi}}, \bibinfo {author} {\bibfnamefont
    {C.-Y.~S.}\ \bibnamefont {Chang}}, \bibinfo {author} {\bibfnamefont {M.-H.}\
    \bibnamefont {Chiu}}, \bibinfo {author} {\bibfnamefont {L.-J.}\ \bibnamefont
    {Li}},\ and\ \bibinfo {author} {\bibfnamefont {F.}~\bibnamefont {Wang}},\
    }\bibfield  {title} {\bibinfo {title} {{Ultrafast generation of
    pseudo-magnetic field for valley excitons in WSe$_2$ monolayers}},\ }\href
    {https://doi.org/10.1126/science.1258122} {\bibfield  {journal} {\bibinfo
    {journal} {Science}\ }\textbf {\bibinfo {volume} {346}},\ \bibinfo {pages}
    {1205} (\bibinfo {year} {2014})}\BibitemShut {NoStop}%
  \bibitem [{\citenamefont {Sie}\ \emph {et~al.}(2015)\citenamefont {Sie},
    \citenamefont {McIver}, \citenamefont {Lee}, \citenamefont {Fu},
    \citenamefont {Kong},\ and\ \citenamefont {Gedik}}]{GedikOpticalStark2015}%
    \BibitemOpen
    \bibfield  {author} {\bibinfo {author} {\bibfnamefont {E.~J.}\ \bibnamefont
    {Sie}}, \bibinfo {author} {\bibfnamefont {J.~W.}\ \bibnamefont {McIver}},
    \bibinfo {author} {\bibfnamefont {Y.-H.}\ \bibnamefont {Lee}}, \bibinfo
    {author} {\bibfnamefont {L.}~\bibnamefont {Fu}}, \bibinfo {author}
    {\bibfnamefont {J.}~\bibnamefont {Kong}},\ and\ \bibinfo {author}
    {\bibfnamefont {N.}~\bibnamefont {Gedik}},\ }\bibfield  {title} {\bibinfo
    {title} {{Valley-selective optical Stark effect in monolayer {WS}$_2$}},\
    }\href@noop {} {\bibfield  {journal} {\bibinfo  {journal} {Nat. Mater.}\
    }\textbf {\bibinfo {volume} {14}},\ \bibinfo {pages} {290} (\bibinfo {year}
    {2015})}\BibitemShut {NoStop}%
  \bibitem [{\citenamefont {Shan}\ \emph {et~al.}(2021)\citenamefont {Shan},
    \citenamefont {Ye}, \citenamefont {Chu}, \citenamefont {Lee}, \citenamefont
    {Park}, \citenamefont {Balents},\ and\ \citenamefont {Hsieh}}]{Hsieh2021nat}%
    \BibitemOpen
    \bibfield  {author} {\bibinfo {author} {\bibfnamefont {J.-Y.}\ \bibnamefont
    {Shan}}, \bibinfo {author} {\bibfnamefont {M.}~\bibnamefont {Ye}}, \bibinfo
    {author} {\bibfnamefont {H.}~\bibnamefont {Chu}}, \bibinfo {author}
    {\bibfnamefont {S.}~\bibnamefont {Lee}}, \bibinfo {author} {\bibfnamefont
    {J.-G.}\ \bibnamefont {Park}}, \bibinfo {author} {\bibfnamefont
    {L.}~\bibnamefont {Balents}},\ and\ \bibinfo {author} {\bibfnamefont
    {D.}~\bibnamefont {Hsieh}},\ }\bibfield  {title} {\bibinfo {title} {{Giant
    modulation of optical nonlinearity by Floquet engineering}},\ }\href
    {https://doi.org/10.1038/s41586-021-04051-8} {\bibfield  {journal} {\bibinfo
    {journal} {Nature}\ }\textbf {\bibinfo {volume} {600}},\ \bibinfo {pages}
    {235} (\bibinfo {year} {2021})}\BibitemShut {NoStop}%
  \bibitem [{\citenamefont {Zhang}\ \emph {et~al.}(2024)\citenamefont {Zhang},
    \citenamefont {Carbin}, \citenamefont {Culver}, \citenamefont {Du},
    \citenamefont {Wang}, \citenamefont {Cheong}, \citenamefont {Roy},\ and\
    \citenamefont {Kogar}}]{Anshul2024natmat}%
    \BibitemOpen
    \bibfield  {author} {\bibinfo {author} {\bibfnamefont {X.}~\bibnamefont
    {Zhang}}, \bibinfo {author} {\bibfnamefont {T.}~\bibnamefont {Carbin}},
    \bibinfo {author} {\bibfnamefont {A.~B.}\ \bibnamefont {Culver}}, \bibinfo
    {author} {\bibfnamefont {K.}~\bibnamefont {Du}}, \bibinfo {author}
    {\bibfnamefont {K.}~\bibnamefont {Wang}}, \bibinfo {author} {\bibfnamefont
    {S.-W.}\ \bibnamefont {Cheong}}, \bibinfo {author} {\bibfnamefont
    {R.}~\bibnamefont {Roy}},\ and\ \bibinfo {author} {\bibfnamefont
    {A.}~\bibnamefont {Kogar}},\ }\bibfield  {title} {\bibinfo {title}
    {{Light-induced electronic polarization in antiferromagnetic Cr$_2$O$_3$}},\
    }\href@noop {} {\bibfield  {journal} {\bibinfo  {journal} {Nat. Mater.}\
    }\textbf {\bibinfo {volume} {23}},\ \bibinfo {pages} {1} (\bibinfo {year}
    {2024})}\BibitemShut {NoStop}%
  \bibitem [{\citenamefont {Uchida}\ \emph {et~al.}(2022)\citenamefont {Uchida},
    \citenamefont {Kusaba}, \citenamefont {Nagai}, \citenamefont {Ikeda},\ and\
    \citenamefont {Tanaka}}]{uchida2022diabatic}%
    \BibitemOpen
    \bibfield  {author} {\bibinfo {author} {\bibfnamefont {K.}~\bibnamefont
    {Uchida}}, \bibinfo {author} {\bibfnamefont {S.}~\bibnamefont {Kusaba}},
    \bibinfo {author} {\bibfnamefont {K.}~\bibnamefont {Nagai}}, \bibinfo
    {author} {\bibfnamefont {T.~N.}\ \bibnamefont {Ikeda}},\ and\ \bibinfo
    {author} {\bibfnamefont {K.}~\bibnamefont {Tanaka}},\ }\bibfield  {title}
    {\bibinfo {title} {{Diabatic and adiabatic transitions between Floquet states
    imprinted in coherent exciton emission in monolayer WSe$_2$}},\ }\href@noop
    {} {\bibfield  {journal} {\bibinfo  {journal} {Sci. Adv.}\ }\textbf {\bibinfo
    {volume} {8}},\ \bibinfo {pages} {eabq7281} (\bibinfo {year}
    {2022})}\BibitemShut {NoStop}%
  \bibitem [{\citenamefont {Kobayashi}\ \emph {et~al.}(2023)\citenamefont
    {Kobayashi}, \citenamefont {Heide}, \citenamefont {Johnson}, \citenamefont
    {Tiwari}, \citenamefont {Liu}, \citenamefont {Reis}, \citenamefont {Heinz},\
    and\ \citenamefont {Ghimire}}]{GhimireNP2023}%
    \BibitemOpen
    \bibfield  {author} {\bibinfo {author} {\bibfnamefont {Y.}~\bibnamefont
    {Kobayashi}}, \bibinfo {author} {\bibfnamefont {C.}~\bibnamefont {Heide}},
    \bibinfo {author} {\bibfnamefont {A.~C.}\ \bibnamefont {Johnson}}, \bibinfo
    {author} {\bibfnamefont {V.}~\bibnamefont {Tiwari}}, \bibinfo {author}
    {\bibfnamefont {F.}~\bibnamefont {Liu}}, \bibinfo {author} {\bibfnamefont
    {D.~A.}\ \bibnamefont {Reis}}, \bibinfo {author} {\bibfnamefont {T.~F.}\
    \bibnamefont {Heinz}},\ and\ \bibinfo {author} {\bibfnamefont
    {S.}~\bibnamefont {Ghimire}},\ }\bibfield  {title} {\bibinfo {title} {Floquet
    engineering of strongly driven excitons in monolayer tungsten disulfide},\
    }\href@noop {} {\bibfield  {journal} {\bibinfo  {journal} {Nat. Phys.}\
    }\textbf {\bibinfo {volume} {19}},\ \bibinfo {pages} {171} (\bibinfo {year}
    {2023})}\BibitemShut {NoStop}%
  \bibitem [{\citenamefont {Park}\ \emph {et~al.}(2022)\citenamefont {Park},
    \citenamefont {Lee}, \citenamefont {Jang}, \citenamefont {Choi},
    \citenamefont {Park}, \citenamefont {Jung}, \citenamefont {Watanabe},
    \citenamefont {Taniguchi}, \citenamefont {Cho},\ and\ \citenamefont
    {Lee}}]{park2022steady}%
    \BibitemOpen
    \bibfield  {author} {\bibinfo {author} {\bibfnamefont {S.}~\bibnamefont
    {Park}}, \bibinfo {author} {\bibfnamefont {W.}~\bibnamefont {Lee}}, \bibinfo
    {author} {\bibfnamefont {S.}~\bibnamefont {Jang}}, \bibinfo {author}
    {\bibfnamefont {Y.-B.}\ \bibnamefont {Choi}}, \bibinfo {author}
    {\bibfnamefont {J.}~\bibnamefont {Park}}, \bibinfo {author} {\bibfnamefont
    {W.}~\bibnamefont {Jung}}, \bibinfo {author} {\bibfnamefont {K.}~\bibnamefont
    {Watanabe}}, \bibinfo {author} {\bibfnamefont {T.}~\bibnamefont {Taniguchi}},
    \bibinfo {author} {\bibfnamefont {G.~Y.}\ \bibnamefont {Cho}},\ and\ \bibinfo
    {author} {\bibfnamefont {G.-H.}\ \bibnamefont {Lee}},\ }\bibfield  {title}
    {\bibinfo {title} {{Steady Floquet--Andreev states in graphene Josephson
    junctions}},\ }\href@noop {} {\bibfield  {journal} {\bibinfo  {journal}
    {Nature}\ }\textbf {\bibinfo {volume} {603}},\ \bibinfo {pages} {421}
    (\bibinfo {year} {2022})}\BibitemShut {NoStop}%
  \bibitem [{\citenamefont {Sobota}\ \emph {et~al.}(2021)\citenamefont {Sobota},
    \citenamefont {He},\ and\ \citenamefont {Shen}}]{sobota2021angle}%
    \BibitemOpen
    \bibfield  {author} {\bibinfo {author} {\bibfnamefont {J.~A.}\ \bibnamefont
    {Sobota}}, \bibinfo {author} {\bibfnamefont {Y.}~\bibnamefont {He}},\ and\
    \bibinfo {author} {\bibfnamefont {Z.-X.}\ \bibnamefont {Shen}},\ }\bibfield
    {title} {\bibinfo {title} {{Angle-resolved photoemission studies of quantum
    materials}},\ }\href@noop {} {\bibfield  {journal} {\bibinfo  {journal} {Rev.
    Mod. Phys.}\ }\textbf {\bibinfo {volume} {93}},\ \bibinfo {pages} {025006}
    (\bibinfo {year} {2021})}\BibitemShut {NoStop}%
  \bibitem [{\citenamefont {Zhang}\ \emph {et~al.}(2022)\citenamefont {Zhang},
    \citenamefont {Pincelli}, \citenamefont {Jozwiak}, \citenamefont {Kondo},
    \citenamefont {Ernstorfer}, \citenamefont {Sato},\ and\ \citenamefont
    {Zhou}}]{ZhangNRMP22}%
    \BibitemOpen
    \bibfield  {author} {\bibinfo {author} {\bibfnamefont {H.}~\bibnamefont
    {Zhang}}, \bibinfo {author} {\bibfnamefont {T.}~\bibnamefont {Pincelli}},
    \bibinfo {author} {\bibfnamefont {C.}~\bibnamefont {Jozwiak}}, \bibinfo
    {author} {\bibfnamefont {T.}~\bibnamefont {Kondo}}, \bibinfo {author}
    {\bibfnamefont {R.}~\bibnamefont {Ernstorfer}}, \bibinfo {author}
    {\bibfnamefont {T.}~\bibnamefont {Sato}},\ and\ \bibinfo {author}
    {\bibfnamefont {S.}~\bibnamefont {Zhou}},\ }\bibfield  {title} {\bibinfo
    {title} {Angle-resolved photoemission spectroscopy},\ }\href
    {https://doi.org/10.1038/s43586-022-00133-7} {\bibfield  {journal} {\bibinfo
    {journal} {Nat. Rev. Methods Primers}\ }\textbf {\bibinfo {volume} {2}},\
    \bibinfo {pages} {54} (\bibinfo {year} {2022})}\BibitemShut {NoStop}%
  \bibitem [{\citenamefont {Boschini}\ \emph {et~al.}(2024)\citenamefont
    {Boschini}, \citenamefont {Zonno},\ and\ \citenamefont
    {Damascelli}}]{boschini2023time}%
    \BibitemOpen
    \bibfield  {author} {\bibinfo {author} {\bibfnamefont {F.}~\bibnamefont
    {Boschini}}, \bibinfo {author} {\bibfnamefont {M.}~\bibnamefont {Zonno}},\
    and\ \bibinfo {author} {\bibfnamefont {A.}~\bibnamefont {Damascelli}},\
    }\bibfield  {title} {\bibinfo {title} {{Time-and Angle-Resolved Photoemission
    Studies of Quantum Materials}},\ }\href@noop {} {\bibfield  {journal}
    {\bibinfo  {journal} {Rev. Mod. Phys.}\ }\textbf {\bibinfo {volume} {96}},\
    \bibinfo {pages} {015003} (\bibinfo {year} {2024})}\BibitemShut {NoStop}%
  \bibitem [{\citenamefont {Kokh}\ \emph {et~al.}(2014)\citenamefont {Kokh},
    \citenamefont {Makarenko}, \citenamefont {Golyashov}, \citenamefont
    {Shegai},\ and\ \citenamefont {Tereshchenko}}]{kokh2014melt}%
    \BibitemOpen
    \bibfield  {author} {\bibinfo {author} {\bibfnamefont {K.}~\bibnamefont
    {Kokh}}, \bibinfo {author} {\bibfnamefont {S.}~\bibnamefont {Makarenko}},
    \bibinfo {author} {\bibfnamefont {V.}~\bibnamefont {Golyashov}}, \bibinfo
    {author} {\bibfnamefont {O.}~\bibnamefont {Shegai}},\ and\ \bibinfo {author}
    {\bibfnamefont {O.}~\bibnamefont {Tereshchenko}},\ }\bibfield  {title}
    {\bibinfo {title} {{Melt growth of bulk Bi$_2$Te$_3$ crystals with a natural
    p--n junction}},\ }\href@noop {} {\bibfield  {journal} {\bibinfo  {journal}
    {CrystEngComm}\ }\textbf {\bibinfo {volume} {16}},\ \bibinfo {pages} {581}
    (\bibinfo {year} {2014})}\BibitemShut {NoStop}%
  \bibitem [{\citenamefont {Chen}\ \emph {et~al.}(2013)\citenamefont {Chen},
    \citenamefont {Xie}, \citenamefont {Feng}, \citenamefont {Yi}, \citenamefont
    {Liang}, \citenamefont {He}, \citenamefont {Mou}, \citenamefont {He},
    \citenamefont {Peng}, \citenamefont {Liu} \emph {et~al.}}]{chen2013tunable}%
    \BibitemOpen
    \bibfield  {author} {\bibinfo {author} {\bibfnamefont {C.}~\bibnamefont
    {Chen}}, \bibinfo {author} {\bibfnamefont {Z.}~\bibnamefont {Xie}}, \bibinfo
    {author} {\bibfnamefont {Y.}~\bibnamefont {Feng}}, \bibinfo {author}
    {\bibfnamefont {H.}~\bibnamefont {Yi}}, \bibinfo {author} {\bibfnamefont
    {A.}~\bibnamefont {Liang}}, \bibinfo {author} {\bibfnamefont
    {S.}~\bibnamefont {He}}, \bibinfo {author} {\bibfnamefont {D.}~\bibnamefont
    {Mou}}, \bibinfo {author} {\bibfnamefont {J.}~\bibnamefont {He}}, \bibinfo
    {author} {\bibfnamefont {Y.}~\bibnamefont {Peng}}, \bibinfo {author}
    {\bibfnamefont {X.}~\bibnamefont {Liu}}, \emph {et~al.},\ }\bibfield  {title}
    {\bibinfo {title} {{{T}unable Dirac fermion dynamics in topological
    insulators}},\ }\href@noop {} {\bibfield  {journal} {\bibinfo  {journal}
    {Sci. Rep.}\ }\textbf {\bibinfo {volume} {3}},\ \bibinfo {pages} {2411}
    (\bibinfo {year} {2013})}\BibitemShut {NoStop}%
  \bibitem [{\citenamefont {Hajlaoui}\ \emph {et~al.}(2012)\citenamefont
    {Hajlaoui}, \citenamefont {Papalazarou}, \citenamefont {Mauchain},
    \citenamefont {Lantz}, \citenamefont {Moisan}, \citenamefont {Boschetto},
    \citenamefont {Jiang}, \citenamefont {Miotkowski}, \citenamefont {Chen},
    \citenamefont {Taleb-Ibrahimi} \emph {et~al.}}]{hajlaoui2012ultrafast}%
    \BibitemOpen
    \bibfield  {author} {\bibinfo {author} {\bibfnamefont {M.}~\bibnamefont
    {Hajlaoui}}, \bibinfo {author} {\bibfnamefont {E.}~\bibnamefont
    {Papalazarou}}, \bibinfo {author} {\bibfnamefont {J.}~\bibnamefont
    {Mauchain}}, \bibinfo {author} {\bibfnamefont {G.}~\bibnamefont {Lantz}},
    \bibinfo {author} {\bibfnamefont {N.}~\bibnamefont {Moisan}}, \bibinfo
    {author} {\bibfnamefont {D.}~\bibnamefont {Boschetto}}, \bibinfo {author}
    {\bibfnamefont {Z.}~\bibnamefont {Jiang}}, \bibinfo {author} {\bibfnamefont
    {I.}~\bibnamefont {Miotkowski}}, \bibinfo {author} {\bibfnamefont
    {Y.}~\bibnamefont {Chen}}, \bibinfo {author} {\bibfnamefont {A.}~\bibnamefont
    {Taleb-Ibrahimi}}, \emph {et~al.},\ }\bibfield  {title} {\bibinfo {title}
    {{Ultrafast surface carrier dynamics in the topological insulator
    Bi$_2$Te$_3$}},\ }\href@noop {} {\bibfield  {journal} {\bibinfo  {journal}
    {Nano Lett.}\ }\textbf {\bibinfo {volume} {12}},\ \bibinfo {pages} {3532}
    (\bibinfo {year} {2012})}\BibitemShut {NoStop}%
  \bibitem [{\citenamefont {S{\'a}nchez-Barriga}\ \emph
    {et~al.}(2017)\citenamefont {S{\'a}nchez-Barriga}, \citenamefont {Battiato},
    \citenamefont {Krivenkov}, \citenamefont {Golias}, \citenamefont
    {Varykhalov}, \citenamefont {Romualdi}, \citenamefont {Yashina},
    \citenamefont {Min{\'a}r}, \citenamefont {Kornilov}, \citenamefont {Ebert}
    \emph {et~al.}}]{sanchez2017subpicosecond}%
    \BibitemOpen
    \bibfield  {author} {\bibinfo {author} {\bibfnamefont {J.}~\bibnamefont
    {S{\'a}nchez-Barriga}}, \bibinfo {author} {\bibfnamefont {M.}~\bibnamefont
    {Battiato}}, \bibinfo {author} {\bibfnamefont {M.}~\bibnamefont {Krivenkov}},
    \bibinfo {author} {\bibfnamefont {E.}~\bibnamefont {Golias}}, \bibinfo
    {author} {\bibfnamefont {A.}~\bibnamefont {Varykhalov}}, \bibinfo {author}
    {\bibfnamefont {A.}~\bibnamefont {Romualdi}}, \bibinfo {author}
    {\bibfnamefont {L.}~\bibnamefont {Yashina}}, \bibinfo {author} {\bibfnamefont
    {J.}~\bibnamefont {Min{\'a}r}}, \bibinfo {author} {\bibfnamefont
    {O.}~\bibnamefont {Kornilov}}, \bibinfo {author} {\bibfnamefont
    {H.}~\bibnamefont {Ebert}}, \emph {et~al.},\ }\bibfield  {title} {\bibinfo
    {title} {{Subpicosecond spin dynamics of excited states in the topological
    insulator Bi$_2$Te$_3$}},\ }\href@noop {} {\bibfield  {journal} {\bibinfo
    {journal} {Phy. Rev. B}\ }\textbf {\bibinfo {volume} {95}},\ \bibinfo {pages}
    {125405} (\bibinfo {year} {2017})}\BibitemShut {NoStop}%
  \bibitem [{ref()}]{ref1}%
    \BibitemOpen
    \href@noop {} {\bibinfo {title} {{See Supplemental Material at [URL] for
    additional information.}}}\BibitemShut {Stop}%
  \bibitem [{\citenamefont {Hajlaoui}\ \emph {et~al.}(2014)\citenamefont
    {Hajlaoui}, \citenamefont {Papalazarou}, \citenamefont {Mauchain},
    \citenamefont {Perfetti}, \citenamefont {Taleb-Ibrahimi}, \citenamefont
    {Navarin}, \citenamefont {Monteverde}, \citenamefont {Auban-Senzier},
    \citenamefont {Pasquier}, \citenamefont {Moisan} \emph
    {et~al.}}]{hajlaoui2014tuning}%
    \BibitemOpen
    \bibfield  {author} {\bibinfo {author} {\bibfnamefont {M.}~\bibnamefont
    {Hajlaoui}}, \bibinfo {author} {\bibfnamefont {E.}~\bibnamefont
    {Papalazarou}}, \bibinfo {author} {\bibfnamefont {J.}~\bibnamefont
    {Mauchain}}, \bibinfo {author} {\bibfnamefont {L.}~\bibnamefont {Perfetti}},
    \bibinfo {author} {\bibfnamefont {A.}~\bibnamefont {Taleb-Ibrahimi}},
    \bibinfo {author} {\bibfnamefont {F.}~\bibnamefont {Navarin}}, \bibinfo
    {author} {\bibfnamefont {M.}~\bibnamefont {Monteverde}}, \bibinfo {author}
    {\bibfnamefont {P.}~\bibnamefont {Auban-Senzier}}, \bibinfo {author}
    {\bibfnamefont {C.}~\bibnamefont {Pasquier}}, \bibinfo {author}
    {\bibfnamefont {N.}~\bibnamefont {Moisan}}, \emph {et~al.},\ }\bibfield
    {title} {\bibinfo {title} {{Tuning a Schottky barrier in a photoexcited
    topological insulator with transient Dirac cone electron-hole asymmetry}},\
    }\href@noop {} {\bibfield  {journal} {\bibinfo  {journal} {Nat. Commun.}\
    }\textbf {\bibinfo {volume} {5}},\ \bibinfo {pages} {3003} (\bibinfo {year}
    {2014})}\BibitemShut {NoStop}%
  \end{thebibliography}
%

\end{document}